\documentclass[aip,preprint]{revtex4-1}

\draft % marks overfull lines with a black rule on the right

\usepackage{graphicx}% Include figure files
\usepackage{amsmath}

\begin{document}

\title{From Close-Packed to Topologically Close-Packed: Formation of Laves Phases in Moderately Polydisperse Hard-Sphere Mixtures} %Title of paper

\author{Beth A. Lindquist}
\affiliation{McKetta Department of Chemical Engineering, University of Texas at Austin, Austin, Texas 78712, USA}
\author{Ryan B. Jadrich}
\affiliation{McKetta Department of Chemical Engineering, University of Texas at Austin, Austin, Texas 78712, USA}
\author{Thomas M. Truskett}
\email{truskett@che.utexas.edu}
\affiliation{McKetta Department of Chemical Engineering, University of Texas at Austin, Austin, Texas 78712, USA}
\affiliation{Department of Physics, University of Texas at Austin, Austin, Texas 78712, USA}

\date{\today}

\begin{abstract}
Particle size polydispersity can help to inhibit crystallization of the hard-sphere fluid into close-packed structures at high packing fractions and thus is often employed to create model glass-forming systems. Nonetheless, it is known that hard-sphere mixtures with modest polydispersity still have ordered ground states. Here, we demonstrate by computer simulation that hard-sphere mixtures with increased polydispersity fractionate on the basis of particle size, and a bimodal subpopulation favors formation of topologically close-packed C14 and C15 Laves phases in coexistence with a disordered phase. The generality of this result is supported by simulations of hard-sphere mixtures with particle-size distributions of four different forms.
\end{abstract}

\pacs{}

\maketitle 

Size-polydisperse hard-sphere (HS) fluids are popular models for computational studies of glass formers. Polydispersity in such systems is well known to inhibit formation of close-packed lattice structures that otherwise readily assemble from the monodisperse fluid upon densification prior to reaching glassy states.~\cite{poly_hs_frac_1,poly_hs_frac_2,poly_hs_frac_3,poly_hs_frac_4,poly_glass_physics_1, poly_glass_physics_2} While polydispersity slows the kinetics of crystallization, fractionation into particle-size-selective close-packed crystalline phases is thermodynamically favored over the fluid in modestly polydisperse mixtures of HS-like particles.~\cite{poly_hs_frac_1,poly_hs_frac_2,poly_hs_frac_3,poly_hs_frac_4} 

Crystallization of fluid mixtures with higher polydispersity, on the other hand, is largely uncharted territory. By analogy to binary HS mixtures, it has been postulated that more complex, non-close-packed crystalline phases could be favored in such systems where the close-packed crystals are penalized.~\cite{poly_hs_frac_1,poly_hs_frac_2,poly_hs_frac_3,poly_hs_frac_4} Analogous behavior is well established for the Dzugutov pair potential, which features a single energetic barrier at distances characteristic of close-packed order, instead favoring self-assembly of dodecagonal quasicrystals.~\cite{dz_potential} Similarly, in a simulation study of a monodisperse HS model with a many-body energy term that favors non-close-packed structure formation, an icosahedra-rich phase is observed--postulated to be either a quasicrystal or of the Frank-Kasper family of crystals (i.e., a quasicrystal approximant).~\cite{close_packed_penalized} 

Motivated by the above, we study the possibility of crystalline assembly in a moderately size disperse HS mixture with particle diameters distributed according to a third-order power law with lower and upper size cutoffs chosen to realize a 12\% polydispersity (i.e., the standard deviation relative to the mean, or the coefficient of variation, $c_{\text{v}}=0.12$).~\cite{poly_glass_physics_1, soft_ternary_laves} Because polydisperse fluids are normally slow to crystallize at high density, we leverage particle-swap Monte Carlo simulations to explore their behavior.~\cite{poly_glass_physics_1, soft_ternary_laves} This method, while powerful, is simple to implement. In addition to standard particle translations, a pair of particles is randomly selected and an attempt is made to swap their diameters. Of the attempted Monte Carlo moves, 20\% are swaps and 80\% are translations. For additional simulation details, see the Appendix.

\begin{figure}[!htb]
  \includegraphics{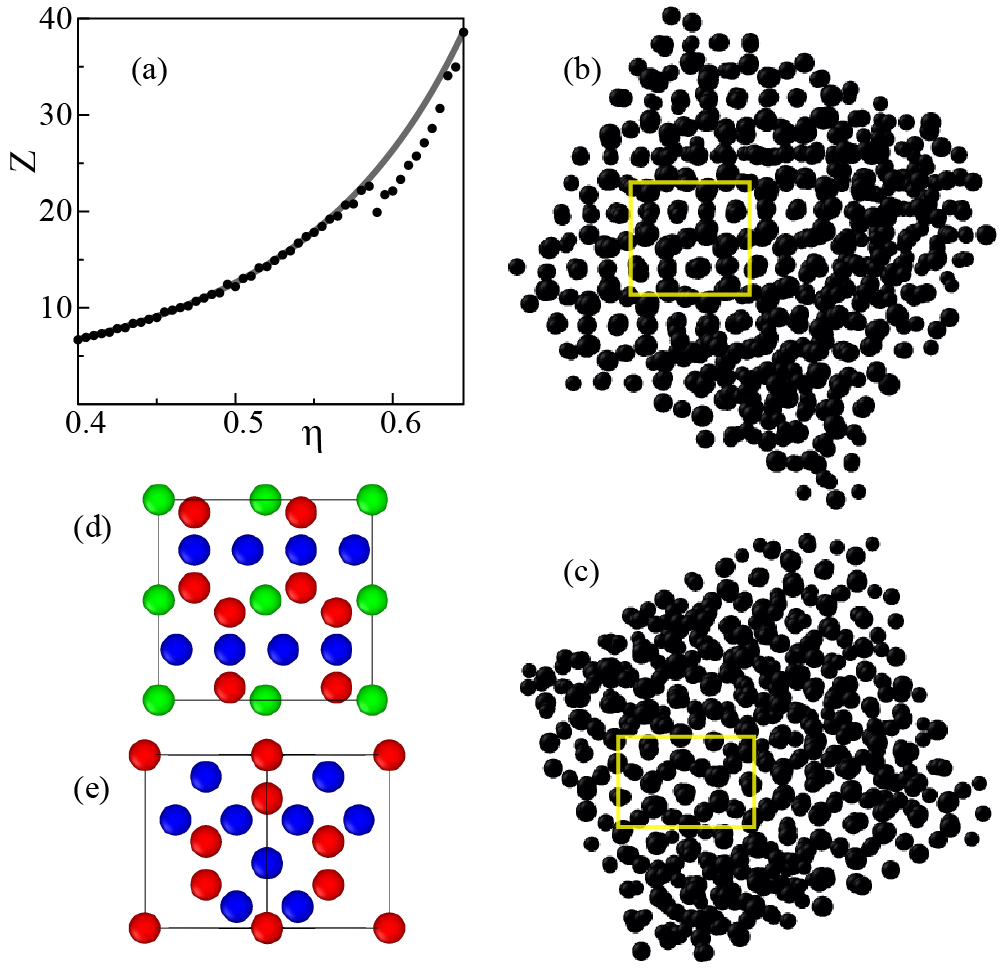}
  \caption{Compressibility factor, $Z \equiv P/\rho k_{\text{B}} T$, as a function of packing fraction, $\eta \equiv 4 \pi \langle r^{3} \rangle \rho / 3$, where $P$ is pressure, $\rho$ is number density, $\langle r^{3} \rangle$ is the third moment of the particle radius distribution, $k_{\text{B}}$ is the Boltzmann constant and $T$ is temperature, for hard spheres with 12\% polydispersity from simulation (symbols) compared to the theoretical prediction (line).~\cite{hs_eos} (b,c) Two representative configurations compressed from $\eta=0.595$ to $\eta=0.66$ from two independent simulations. (d,e) Visualizations of a repeating unit for the C14 and C15 lattices, respectively. The C14 repeating unit shown is an orthogonalization of the hexagonal C14 unit cell, and the cubic C15 unit cell has been rotated by 45$^\circ$. Throughout the manuscript, particles are depicted with diameters that are 50\% of their true diameter for clarity.}
  \label{fgr:Zveta} %add yellow boxes
\end{figure}

Upon densification, the pressure of the aforementioned HS mixture noticeably deviates from the theoretical prediction;~\cite{hs_eos} see Fig.~\ref{fgr:Zveta}a. Visualization of configuration data after the abrupt drop in pressure reveals the presence of large ordered domains; see Fig.~\ref{fgr:Zveta}b,c for two snapshots from independent simulations, where particle diameters are scaled by 0.5 in all snapshots for visual clarity. The crystalline domains are reminiscent of Laves phases--a class of topologically close-packed structures that are known to be thermodynamically stable for certain binary mixtures of hard spheres.~\cite{frank_kasper_and_laves} Specifically, comparison of the structures outlined in yellow in Fig.~\ref{fgr:Zveta}b,c to ideal repeating units of the C14 and C15 lattices in Fig.~\ref{fgr:Zveta}d,e reveals that the former are consistent with the latter, with some substitutional disorder present due to the polydispersity.

\begin{figure}[!htb]
  \includegraphics{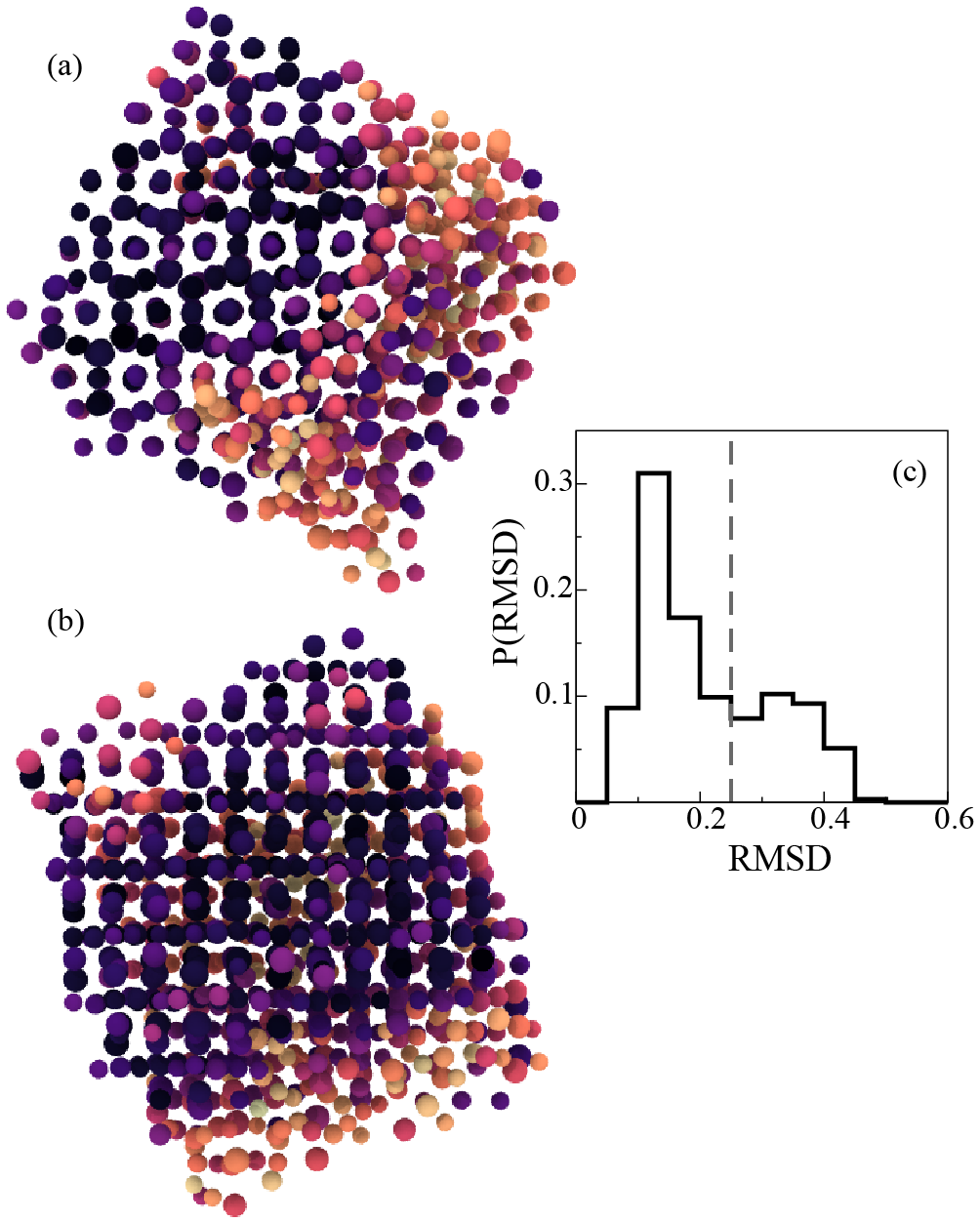}
  \caption{(a-b) Two views of the configuration from Fig.~1b, color-coded by the RMSD metric described in the main text and in the Appendix. Darker regions indicate C14 crystalline domains and lighter regions appear disordered. (c) Probability distribution as a function of RMSD for the snapshot in panel a,b.}  
  \label{fgr:RMSD}
\end{figure}

To confirm the presence of Laves phases, we applied a template-matching scheme for the C14 and C15 lattices using the lattice constants extracted from the simulation box above, computing the root-mean-square deviation (RMSD) from an idealized lattice; see the Appendix for details. Using the C14 template, the lattice shown in Fig.~\ref{fgr:Zveta}b above is color-coded according to the RMSD in Fig.~\ref{fgr:RMSD}a,b, where co-existence of fluid and crystalline regions on the basis of color is clear. The distribution of RMSDs, shown in Fig.~\ref{fgr:RMSD}c, is weakly bimodal, functioning as an order metric to segregate the particles into crystalline and fluid regions--though the interface between the fluid and crystal regimes weakens the biomodality. We chose a value of 0.25 as the maximum RMSD for a crystalline particle in the analysis that follows. A parallel analysis for the configuration shown in Fig.~\ref{fgr:Zveta}c is given in the Appendix. 

\begin{figure}[!htb]
  \includegraphics{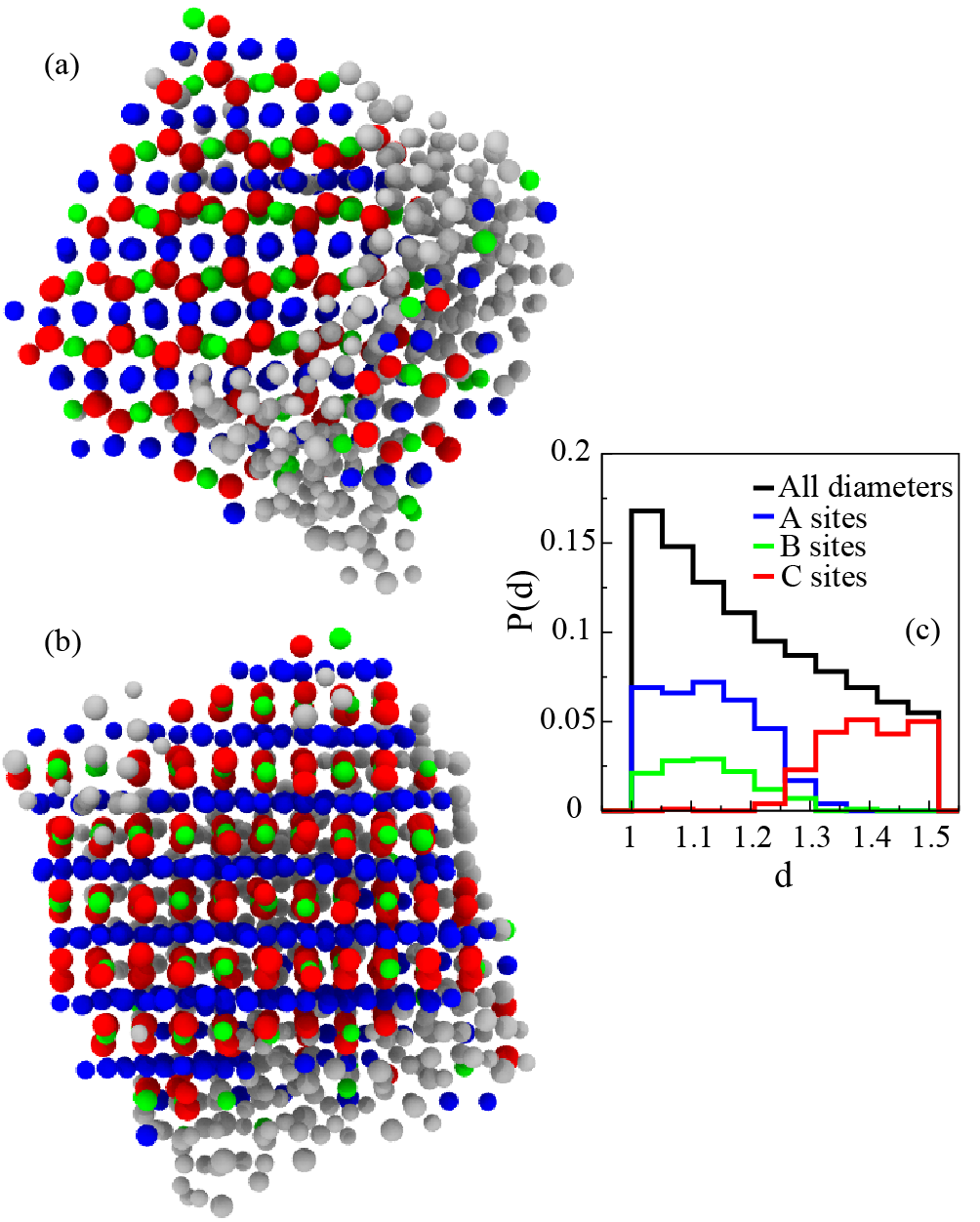}
  \caption{(a,b) The same configuration shown in Fig.~2a,b color-coded by particle type. A sites are blue, B sites are green, and C sites are red. (c) The overall probability distribution for the diameters (black), and the probability distributions for the A, B, and C sites; all distributions are normalized by the total number of particles in the simulation box.}
  \label{fgr:latticeSites}
\end{figure}

As shown in Fig.~\ref{fgr:Zveta}d,e, the C14 and C15 lattices have three and two distinct lattice positions, respectively; the template-matching scheme above allows us to label the crystalline particles by lattice site; see Fig.~\ref{fgr:latticeSites}a,b. Because 1) the C14 and C15 lattices are closely related structurally~\cite{binary_laves_1} and 2) the system possesses significant substitutional disorder, the template-matching scheme cannot always distinguish between the two lattices; however, labeling the lattice sites can be helpful to visualize C14 and C15 motifs. For instance, the few C15 domains co-existing with the larger C14 domain can be visualized where the green particles are vertically offset as shown in Fig.~\ref{fgr:latticeSites}a. The co-existence of C14 and C15 structures is not surprising; it is known that the Laves phases have very similar free energies for certain bidisperse mixtures.~\cite{binary_laves_1}

Similar to prior work for systems with lower polydispersity,~\cite{poly_hs_frac_1,poly_hs_frac_2,poly_hs_frac_3,poly_hs_frac_4} we observe fractionation on the basis of particle diameter. Fig.~\ref{fgr:latticeSites}c shows a histogram of the distributions of diameters associated with distinct lattice sites in comparison with the complete distribution of diameters in the mixture (black line). The subset of smaller diameters nearly exclusively occupy the A (blue) and B (green) lattice sites, whereas the larger particles occupy the C (red) lattice sites. (The A and B lattice sites have a highly similar coordination shell structure in the C14 lattice.) 

While the preceding self-assembled structures do not prove that the Laves phases are thermodynamically stable for the above polydisperse mixture, the characteristics of the crystalline phase are consistent with the known phase behavior of binary HS systems. Based on free energy calculations, Laves phases are expected if the diameter ratio of large to small particles ($d_{\text{L}}/d_{\text{S}}$) falls within the range 1.19-1.35.~\cite{binary_laves_1} This range is in direct accord with the approximate size ratio we find in the solid phase: $d_{\text{L}}/d_{\text{S}}\approx1.24$ (where $d_{\text{L}}$ and $d_{\text{S}}$ are the averages of the C site and A+B site distributions, respectively). The existence of Laves phases for similar size ratios has also been confirmed experimentally.~\cite{exp_binary_1} In essence, the larger range of particle diameters characteristic of higher polydispersity allows for thermodynamically driven size-segregation resulting in domains that can form complex lattices.

\begin{figure}[!htb]
  \includegraphics{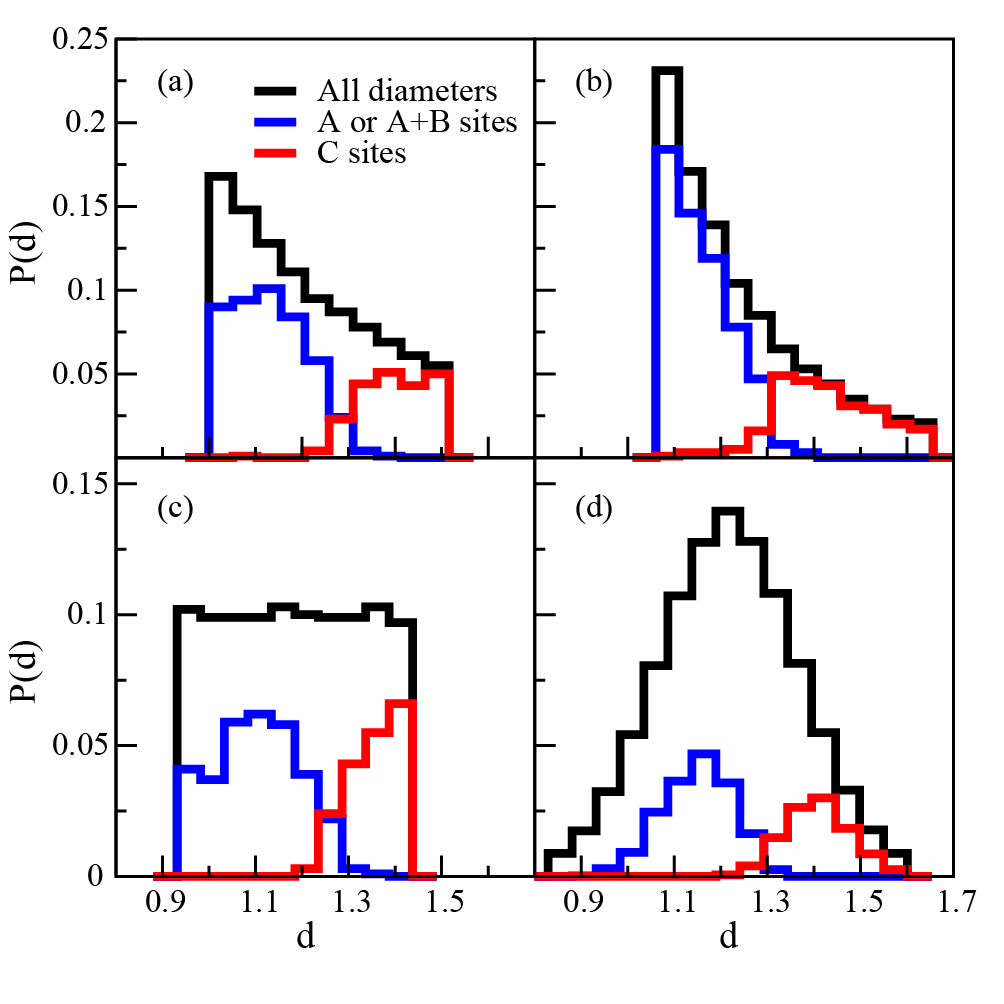}
  \caption{(a-d) For the third-order, sixth-order, uniform, and Gaussian distributions, respectively, the A (or A+B) distribution (blue, smaller diameters), the C distribution (red, larger diameters), and the the total distribution (black). Where the C14 template was used, the A and B particles have been combined into the blue distribution. All distributions are normalized by the total number of particles in the simulation box.}
  \label{fgr:extra_dist}
\end{figure}

%other polydispersities
To evaluate the generality of Laves phase formation, we explore other particle size distributions with the same mean diameter and coefficient of variation ($c_{\text{v}}$). Specifically, we repeated the above simulation analysis for uniform, sixth-order power law, and Gaussian size distribution forms--each of which are shown in Fig.~\ref{fgr:extra_dist}a-d in conjunction with the third-order distribution studied above. The first two distributions are effectively flattened and stretched versions, respectively, of the third-order distribution. The Gaussian distribution in Fig.~\ref{fgr:extra_dist}d provides a qualitatively distinct scenario, where particle diameters are concentrated around the mean of the distribution. Nonetheless, in all mixtures we find the presence of Laves phases, driven by size-selective fractionation into solid and fluid phases. The template-matching scheme finds that bimodal distributions with respect to diameter accommodate formation of the Laves phases, where the smaller particles occupy the A (or A+B) sites and the larger particles occupy the C sites. While the shapes of the sub-distributions are sensitive to the form of the total distribution, $d_{\text{L}}/d_{\text{S}}$ is between 1.22 and 1.24 for all four cases. In binary mixtures, $d_{\text{L}}/d_{\text{S}}=1.225$ is the size ratio where the Laves phases have a maximum packing fraction.~\cite{binary_laves_1}

A 2(A+B):1C stoichiometric ratio is needed to form the C14 lattice; similarly, the required ratio is 2A:1C for the C15 lattice. Therefore, the distributions that are shifted towards greater numbers of smaller particles are particularly well suited to form Laves phases. Indeed, the Laves phase domains present in the the sixth-order power law distribution comprise 85\% of the total particles in the simulation. By contrast, for the case of the flat-top distribution, the Laves phase regime identified by template-matching only accounts for 51\% of the total particles in the simulation. 

One limitation of our simulation protocol can be gleaned from the size distributions shown in Fig.~\ref{fgr:extra_dist}, where there are cases for which all (or almost all) of the particles of a given size have been incorporated into the crystal structure. As a result, it is possible that the crystalline phase has incorporated particles with a non-ideal diameter compared to a system connected to an infinite reservoir of particles. For this reason, it is known that finite-size effects are particularly relevant to mixtures.~\cite{poly_hs_frac_finite} On the other hand, the relative insensitivity of $d_{\text{L}}/d_{\text{S}}$ with respect to the form of the distribution and the consistency of Laves phase formation is suggestive that these two findings are not especially sensitive to finite-size effects. Furthermore, we have performed an identical analysis to that shown in Figs.~\ref{fgr:RMSD} and~\ref{fgr:latticeSites} for $N=10000$ and found near quantitative agreement between the two system sizes; see the Appendix. Ultimately though, simulations in the isobaric-semi-grand canonical ensemble could be useful to confirm the reported properties of the crystalline phase.~\cite{poly_hs_frac_finite}       

The above results may be useful in the interpretation of several prior studies. One such study employing polydisperse soft spheres with a $c_{\text{V}}\approx0.24$ found a high-density inhomogeneous state comprised of coexisting ordered and fluid phases with strong particle size segregation.~\cite{soft_poly_laves_1,soft_poly_laves_2} The ordered phase was not identified; however, the published snapshots are consistent with a Laves phase, and $d_{\text{L}}/d_{\text{S}}\approx1.26$ for the ordered phase.~\cite{soft_poly_laves_2} In another study, a soft sphere ternary mixture yielded a structurally similar solid phase dominated by two of the three particle types with $d_{\text{L}}/d_{\text{S}}=1.25$.~\cite{soft_ternary_laves} Taken together, it appears that the formation of Laves phases in HS systems may be a far more common and general phenomenon than previously recognized. Therefore, studies of moderately polydisperse HS fluids should explicitly consider the the possibility of Laves phase formation. 

Furthermore, the binary HS phase diagram may provide a generally useful reference to anticipate and interpret the crystalline phases that emerge in polydisperse mixtures. A recent study of a polydisperse HS system with $c_{\text{V}}\approx0.23$ observed partial crystallization of the aluminum diboride (AlB$_{2}$) lattice with $d_{\text{L}}/d_{\text{S}}\approx2$, a value within the AlB$_{2}$ stability window for the binary HS system.~\cite{poly_hs_frac_AlB2} Finally, we note that the disordered phase that exists in coexistence with the Laves phases may have interesting properties (e.g., it may be a particularly good glass former)--an interesting avenue for future study. 

\begin{acknowledgments}
This research was partially supported the Welch Foundation (F-1696) and by the National Science Foundation through the Center for Dynamics and Control of Materials: an NSF MRSEC under Cooperative Agreement No. DMR-1720595. We acknowledge the Texas Advanced Computing Center (TACC) at The University of Texas at Austin for providing HPC resources.
\end{acknowledgments}

\setcounter{figure}{0}
\renewcommand\thefigure{A\arabic{figure}}
\renewcommand{\thesection}{\thepart .\arabic{section}}

\section*{Appendix}

\subsection{Additional Simulation Protocol}

All simulations employed a finite $N$ approximation to the various sphere diameter ($d$) distributions, where $N$ is the number of particles in the simulation. Practically, each probability distribution $P(d)$ was divided into $N$ contiguous, equi-probability chunks with the midpoints taken as the particle diameters. 

With the above finite population approximation scheme, we performed standard NVT Monte Carlo simulations with the addition of particle swap moves. Translation and swap moves were selected at random with an $80:20$ weighting. Translation involved randomly selecting a particle and attempting a random translation with a maximum displacement chosen to achieve an average acceptance rate between $17-25\%$. In the case of a swap move, two particles were chosen at random, and their diameters were exchanged if no overlaps were generated by the swap. For the power law and flat top distributions, $N=1000$ was employed, while $5000$ particles were used for the Gaussian distribution to provide adequate sampling of the tails of the distribution. A simulation with $N=100$ was used for the pressure calculation in Fig. 1a of the main text--useful primarily to identify the approximate density at which freezing takes place ($\eta=0.58-0.6$).

Initial configurations were generated via a random sequential addition protocol at $\eta=0.25$ and quickly compressed to $\eta=0.595$. Simulations were run until the sample crystallized, after which swap moves were disabled and the system was quickly compressed to $\eta=0.66$ for the flat top and power law distributions and $\eta=0.65$ for the Gaussian, at which point the latter mixture jammed and could not be compressed further. The final compression was utilized to help lock the particles into their idealized lattice positions for identification purposes. Only very small scale particle movements occurred during this step.

\subsection{Description of Template-Matching Procedure}

For the template-matching procedure referenced in the main text, we first visually analyze a configuration to find an orientation that displays the 10-fold coordination characteristic of Laves phases such as those shown in Fig. 1b,c of the main text. We rotate the configuration so that a large ordered domain is oriented properly with respect to the templates shown in Fig. 1d,e. For each particle in the simulation box, we assign it to an ideal lattice position in the template and then search for neighboring particles in the simulation that are closest to the other ideal lattice positions prescribed by the template, all while using periodic boundary conditions consistent with the original simulation box. The C14 template has 41 particles; the C14 unit cell only has 12 particles, but we have orthogonalized the unit cell (which doubles the number of particles) and replicated particles that fall on the edges and corners as dictated by the periodicity of the lattice. The C15 template has 34 particles after replicating the particles on the corners and edges of the 24 particle unit cell. The lattice constants for the template are determined by averaging over several approximate realizations of the template in the simulation box.

We calculate the root-mean-squared deviation (RMSD) of the local environment of the selected particle in the simulation box with respect to the template for every lattice position in the template, using the minimum image convention to place the selected particle as close to the center of the template as possible. The reported RMSD value for a selected particle is given by the lowest value of the RMSD over all of the template sites, and its lattice site assignment corresponds to the template site that has the lowest RMSD as well. Therefore, the assignments based on the template-matching scheme are only interdependent insofar as they are all calculated from the same structure; there is no check for consistency of the lattice type assignment among groups of particles. The largely correct arrangements of particle types observed in a large domain of Fig. 3a with respect to an ideal C14 lattice is a consequence of the reasonable accuracy of the template-matching scheme, and, as described in the main text, deviations from the ideal arrangement can indicate mixing of other Laves phases (C15 in the case of Fig. 3a). They could also indicate a local defect in the crystalline region. 

Because the template-matching is sensitive to the orientation of the simulation box, it is not guaranteed to be an exhaustive characterization of all of the particles in the system. If the large majority of the particles form a single ordered domain (as seems to be the case in Fig. 3a), then the scheme should provide a reasonable description of all of the particles. However, if multiple domains exist that are not oriented consistently with one another, it is possible that the template-matching scheme will only accurately describe some of the domains. Therefore, we intend that the template-matching scheme is a tool to characterize the composition of large ordered domains that can be visualized within a simulation box, and not as a fully automated and exhaustive means to search for Laves phases in any arbitrary system. 

\subsection{Analysis of Template-Matching for Additional Structure}

\begin{figure}[!htb]
  \includegraphics{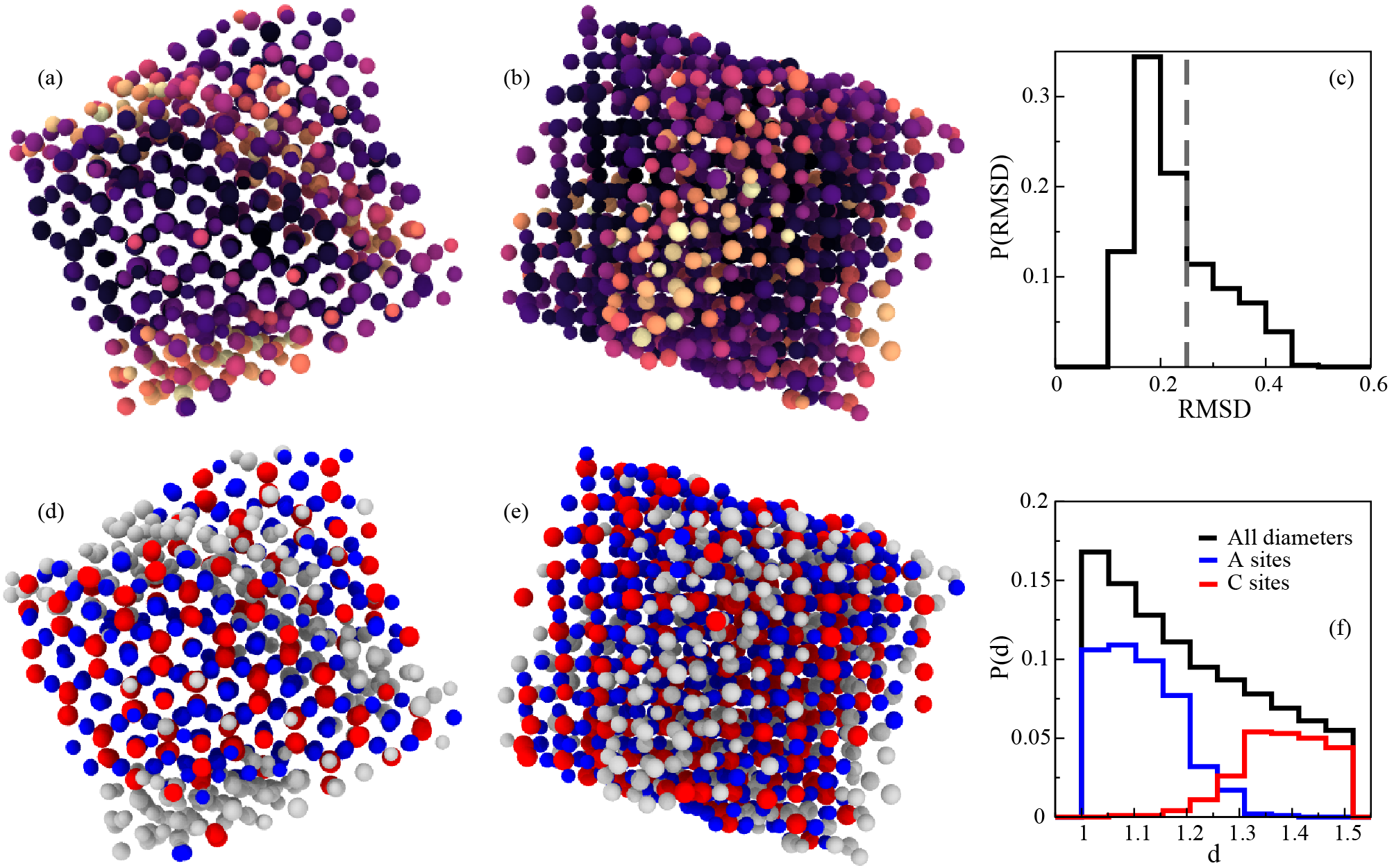}
  \caption{(a-b) Two views of the configuration shown in Fig. 1c of the main text, color-coded by the RMSD metric described above. Darker regions indicate C15 crystalline domains. (c) Probability distribution as a function of RMSD for the snapshot in panel a,b. (d-e) The same configuration shown in panels a and b, color-coded by particle type. A sites are blue, and C sites are red. (f) The overall probability distribution for the diameters (black), and the probability distributions for the A and C sites. All distributions are normalized by the total number of particles in the simulation box.}  
  \label{fgr:SIp3}
\end{figure}

Here, we examine the structure shown in Fig. 1c of the main text. Visually, it seems that the ordered domain in Fig. 1c is largely composed of C15 motifs. Therefore, we performed the analysis below with the C15 template instead of the C14 template. In Fig.~\ref{fgr:SIp3}a,b, see the configuration color-coded such that darker regions indicate C15 crystalline regions (low RMSD values) and lighter regions appear disordered (high RMSD values). Fig.~\ref{fgr:SIp3}c gives the distributions of RMSD values for the particles, and, as in the main text, a cut-off of 0.25 was used as the upper bound for the Laves crystalline region. 

The configuration, color-coded by particle type, is shown in Fig.~\ref{fgr:SIp3}d,e. Visually it is possible to discern that the red (C site) particles tend to be larger than the blue (A site) particles. This trend is born out in the particle diameter distributions shown in Fig.~\ref{fgr:SIp3}f for the A, C, and total particle size distributions. Identically with the results in the main text, $d_{\text{L}}/d_{\text{S}}=1.24$. 

\subsection{Visualizations of Structures Corresponding to Other Size Distributions}

\begin{figure}[!htb]
  \includegraphics{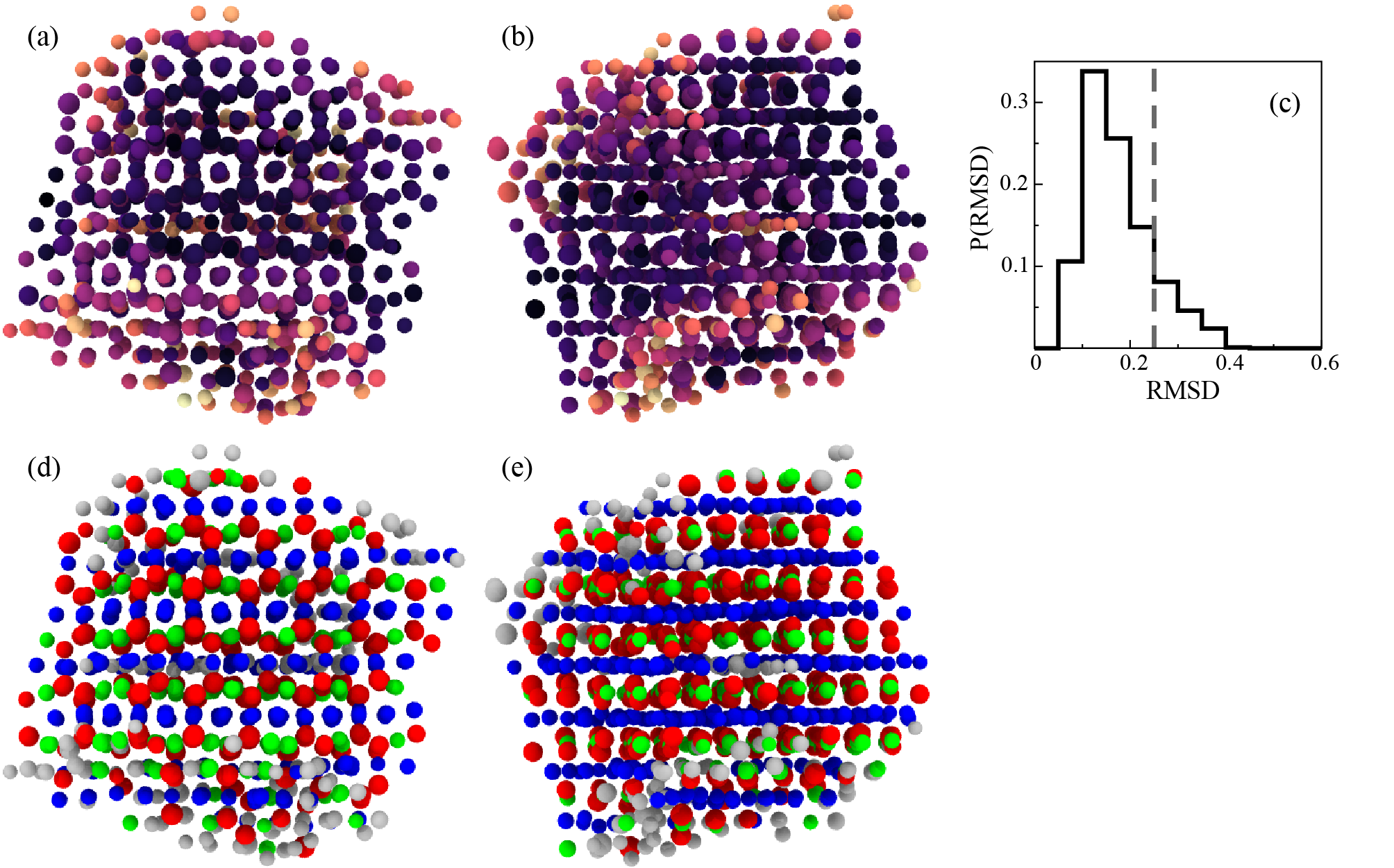}
  \caption{(a-b) Two views of the final configuration from the sixth-order power law distribution, color-coded by the RMSD metric described above. Darker regions indicate Laves phase crystalline domains. (c) Probability distribution as a function of RMSD for the snapshot in panel a,b. (d-e) The same configuration shown in panels a and b, color-coded by particle type. A sites are blue, B sites are green, and C sites are red.}  
  \label{fgr:SIp6}
\end{figure}

\begin{figure}[!htb]
  \includegraphics{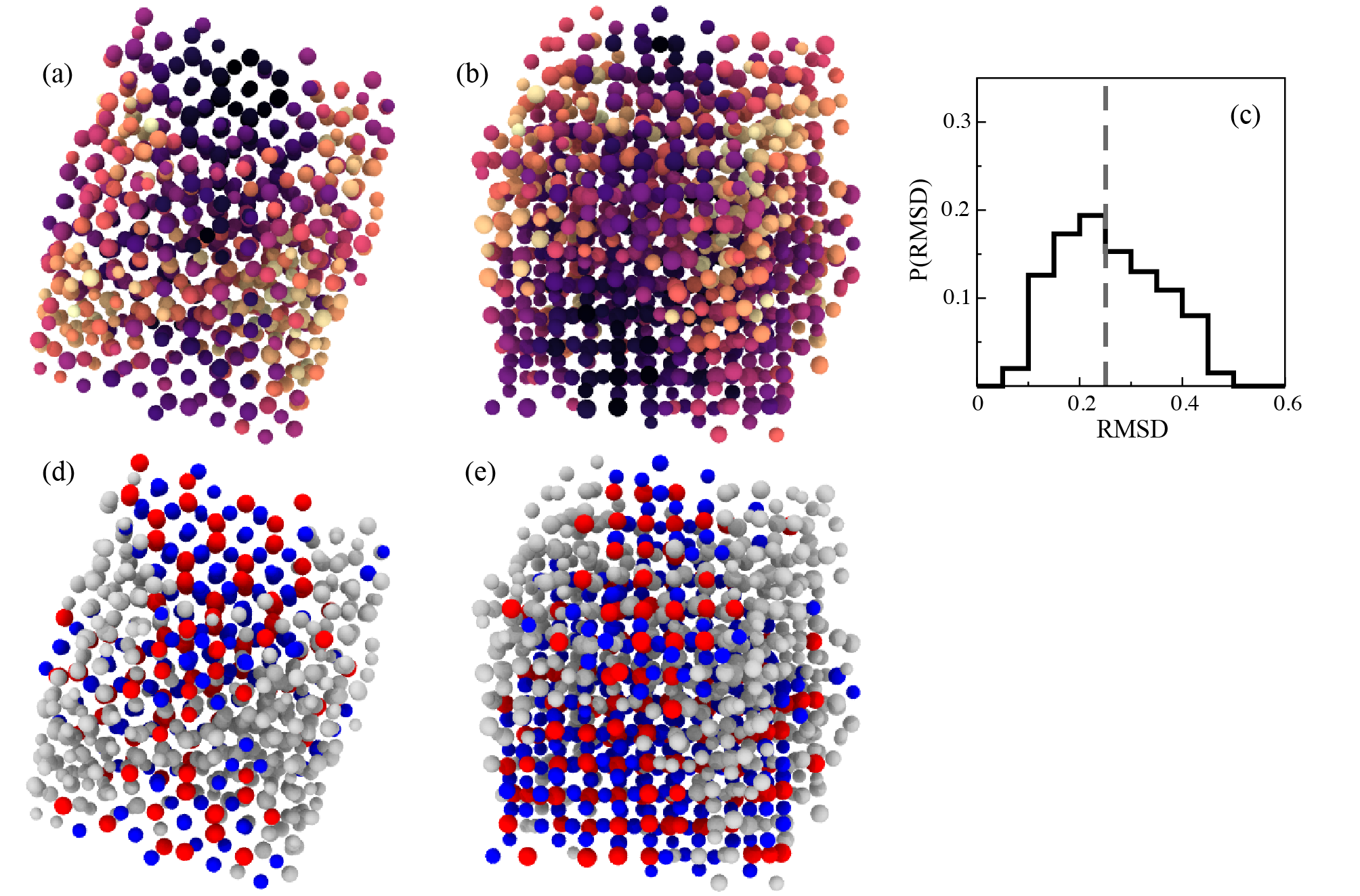}
  \caption{(a-b) Two views of the final configuration from the uniform distribution, color-coded by the RMSD metric described above. Darker regions indicate Laves phase crystalline domains. (c) Probability distribution as a function of RMSD for the snapshot in panel a,b. (d-e) The same configuration shown in panels a and b, color-coded by particle type. A sites are blue, and C sites are red.}
  \label{fgr:SIft}
\end{figure}

\begin{figure}[!htb]
  \includegraphics{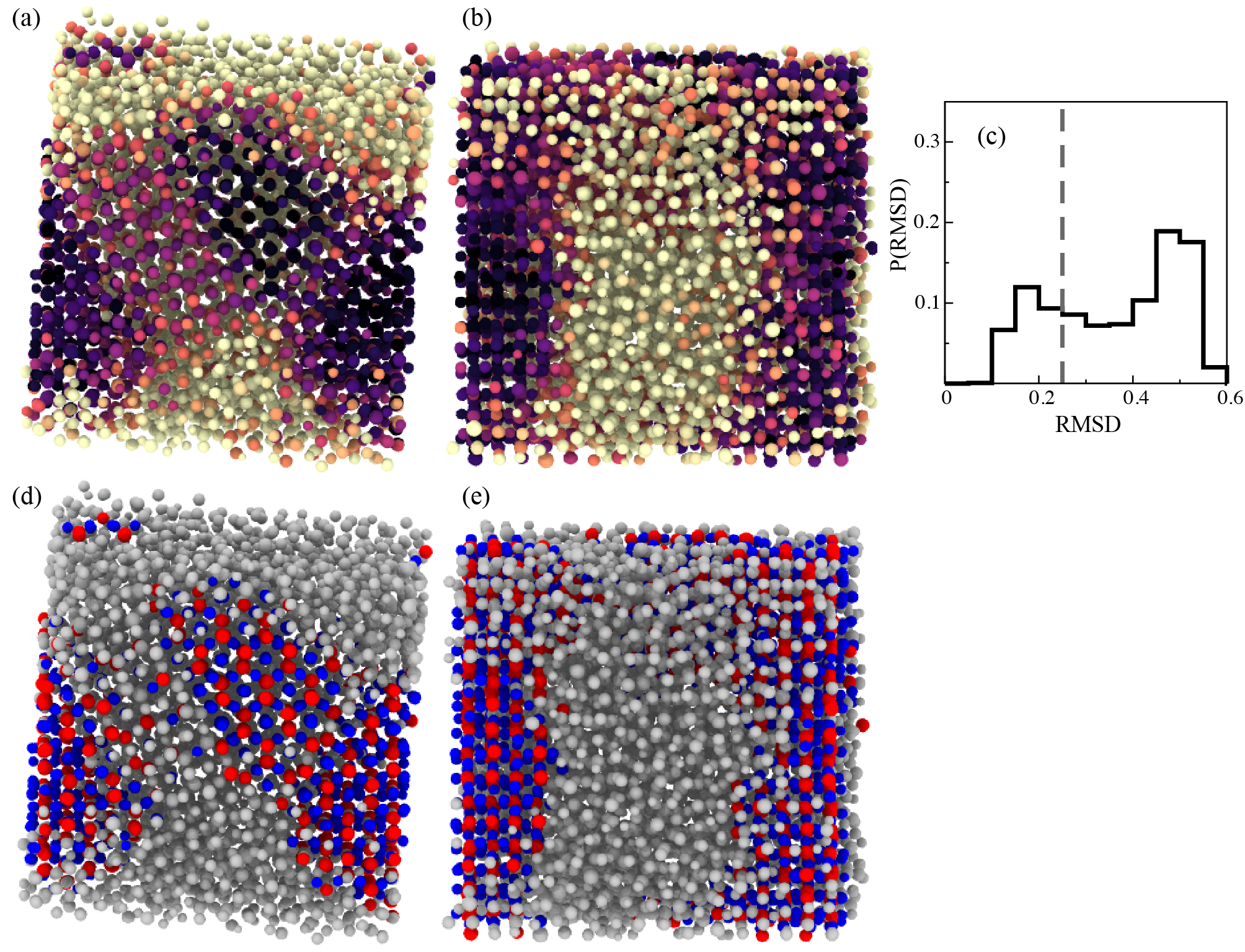}
  \caption{(a-b) Two views of the final configuration from the Gaussian distribution, color-coded by the RMSD metric described above. Darker regions indicate Laves phase crystalline domains. (c) Probability distribution as a function of RMSD for the snapshot in panel a,b. (d-e) The same configuration shown in panels a and b, color-coded by particle type. A sites are blue, and C sites are red.}
  \label{fgr:SIgauss}
\end{figure}

In Fig. 4b-d of the main text, we reported the particle size distributions for the A (or A+B) and C sites for a sixth-order power law, a uniform, and a Gaussian distribution, respectively. In Figs.~\ref{fgr:SIp6}-~\ref{fgr:SIgauss}, we show the corresponding structures, color-coded by both RMSD and particle type, as well as the probability distribution of the former. For the sixth-order power law, the C14 template was used and for the uniform and Gaussian distributions, the C15 template was used. The respective values for $d_{\text{L}}/d_{\text{S}}$ were: 1.23, 1.22, and 1.23.

As mentioned above, we found that we needed to use a larger simulation box ($N=5000$) for the Gaussian distribution. Since the particles that comprise the Laves phases are generally selected from the ends of the distributions, the Gaussian distribution furnished fewer particles of the appropriate size ratio to form Laves phase. For simulations using $N=1000$, some degree of ordering was apparent, but larger Laves phase domain were not visible. In the $N=5000$ simulations, some Laves phase domains self-assembled as can be seen in Fig. ~\ref{fgr:SIgauss}, possibly in co-existence with other crystalline motifs--an interesting avenue for future work. 

\clearpage

\subsection{Evaluation of Finite-Size Effects}

To test for the importance of finite-size effects, we performed an identical analysis as that shown in Figs. 2--3 of the main text after increasing $N$ from 1000 to 10000. Laves phase formation occurred in both simulations, and the size distributions were highly similar. We observed quantitative agreement of $d_{\text{L}}/d_{\text{S}}\approx1.24$ in both cases.

\begin{figure}[!htb]
  \includegraphics[height=13cm,keepaspectratio]{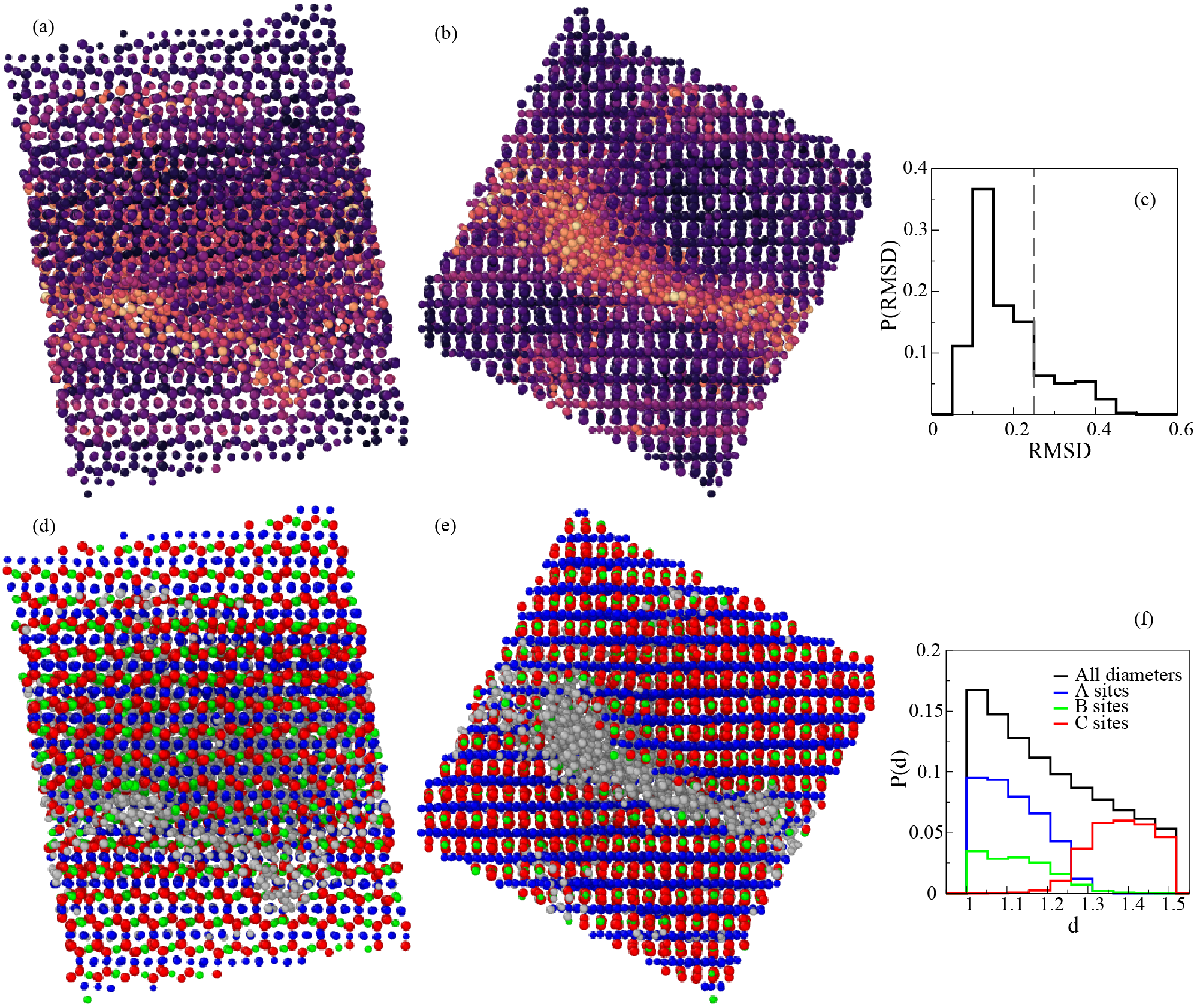}
  \caption{(a-b) Two views of the final configuration from the third order power law distributions where $N=10000$, color-coded by the RMSD metric described above. Darker regions indicate Laves phase crystalline domains. (c) Probability distribution as a function of RMSD for the snapshot in panel a,b. (d-e) The same configuration shown in panels a and b, color-coded by particle type. A sites are blue, B sites are green, and C sites are red. (f) The overall probability distribution for the diameters (black), and the probability distributions for the A, B, and C sites.}
  \label{fgr:SI-N10000}
\end{figure}

\clearpage

%\bibliography{bibliography.bib}


\end{document}